\documentclass[draftclsnofoot,onecolumn, 12pt]{IEEEtran}
\usepackage{color}
\usepackage{graphicx, pstool}
\usepackage{amsmath}
\usepackage{amsfonts}
\usepackage{amssymb}
\usepackage{booktabs}
\usepackage{dsfont}
\usepackage{cases}
\usepackage{tabularx}
\usepackage{boxedminipage}
\usepackage{url}
\usepackage{mathrsfs}
\usepackage{cite}
\usepackage{authblk}
\usepackage{amsmath}
\usepackage{setspace}
\usepackage{mathtools}
\usepackage{breqn}
\usepackage{algorithm}
\usepackage[noend]{algpseudocode}
\usepackage[english]{babel} 
\usepackage{bm} 
\usepackage{mathrsfs}

\usepackage{algpseudocode} 
\algnewcommand{\algorithmicgoto}{\textbf{go to}}%
\algnewcommand{\Goto}[1]{\algorithmicgoto~\ref{#1}}%
\newtheorem{theorem}{Theorem}
\newtheorem{remark}{Remark}

\usepackage{eurosym}
\begin{document}

\title{ Optimal Centralized Dynamic-Time-Division-Duplex}
\author{Mohsen Mohammadkhani Razlighi, Nikola Zlatanov, Shiva Raj Pokhrel, and Petar Popovski
\thanks{This work has been published in part at  IEEE WCNC 2019 \cite{8885925}.}
\thanks{M. M. Razlighi  N. Zlatanov are  with the Department of Electrical and Computer Systems Engineering, Monash University, Melbourne, VIC 3800, Australia (e-mails:  mohsen.mohammadkhanirazlighi@monash.edu and  nikola.zlatanov@monash.edu.)

S. R. Pokhrel is with the school of IT, Deakin University, Geelong Australia (e-mail: shiva.pokhrel@deeakin.edu.au)

P. Popovski is with the Department of Electronic Systems, Aalborg University, Denmark (e-mail: petarp@es.aau.dk).
}
}
\maketitle

\begin{abstract}
In this paper, we derive the optimal  centralized dynamic-time-division-duplex (D-TDD)  scheme  for a wireless network comprised of $K$  full-duplex nodes impaired by self-interference and additive white Gaussian noise. As a special case, we also provide the optimal centralized D-TDD scheme when the nodes are half-duplex as well as when the wireless network is comprised of both half-duplex and full-duplex nodes. Thereby, we derive the  optimal adaptive scheduling of the reception, transmission, simultaneous reception and transmission,  and silence at every node in the network in each time slot such that the rate region of the network is maximized. The performance of the optimal centralized D-TDD can serve as an upper-bound to any other TDD scheme, which is useful in qualifying the relative performance of TDD schemes. The numerical results show that the proposed  centralized D-TDD  scheme achieves significant rate gains over existing centralized D-TDD schemes.

\end{abstract}

\section{Introduction}\label{Sec-Intro}
Time-division duplex (TDD) is a communication protocol where the receptions and transmissions of the network nodes are allocated to non-overlapping  time slots in the same frequency band. TDD has wide use in  3G, 4G, and 5G since it  allows for an easy and flexible control over the flow of uplink and downlink data   at the nodes, which is achieved by changing the portion of time slots allocated to reception and transmission at the nodes \cite{holma2011lte,TechNoteLET}.

In general, the TDD scheme can be  static or dynamic. In static-TDD, each node pre-allocates a fraction of the total number of time slots  for transmission and the rest of the time slots  for reception regardless of the channel conditions and the interference in the network  \cite{holma2011lte}. Due to the scheme being static, the time slots in which the nodes perform reception and the time slots in which the nodes perform transmission are  prefixed and unchangeable over long periods \cite{TechNoteLET}. On the other hand, in dynamic (D)-TDD, each time slot can be dynamically allocated either for reception or  for transmission at the  nodes based on the channel gains of  the network links in order to  maximize the  overall network performance. Thereby, D-TDD schemes achieve higher performance gain compared to  static-TDD schemes at the expense of overheads. As a result, D-TDD schemes have attracted significant research interest, see \cite{8408762,8016428,7386709,957300,8119931,1044604,7073589,8399860} and references therein. Motivated by this, in this paper we investigate D-TDD schemes.
 

D-TDD schemes can be implemented in either distributed or centralized fashion. In distributed D-TDD schemes, the individual nodes, or a group of nodes, make  decisions for transmission, reception, or silence  without  synchronizing with the rest of the nodes in the network \cite{8403595,8334580,8812711,6213034}. As a result, a distributed D-TDD scheme is practical for implementation, however, it does not maximize the overall network performance. On the other hand, in centralized D-TDD schemes, the decision of whether a node should receive, transmit or stay silent in a given time slot is performed at a central processor in the network, which then informs the node about its decision. To this end, centralized D-TDD schemes require full channel state information (CSI) of all network links at the central processor. In this way, the receptions, transmissions, and silences of the  nodes are synchronized by the central processor  in order to   maximize the overall network performance. Since centralized D-TDD schemes require full CSI of all network links, they induce excessive overhead and thus are not practical for implementation. However, knowing the performance of the optimal centralized D-TDD scheme is highly valuable since it serves as an upper bound and thus serves as an (unattainable) benchmark for any practical TDD scheme. The optimal centralized D-TDD scheme for a wireless network is an  open problem. Motivated by this, in this paper we derive the optimal centralized D-TDD scheme for a wireless network.

A network node can operate in two different modes, namely, full-duplex (FD)  mode and  half-duplex (HD) mode. In the FD mode, transmission and reception at the node can occur simultaneously and in the same frequency band. However, due to the in-band simultaneous reception and transmission, nodes are impaired by self-interference (SI), which occurs due to leakage of energy from the transmitter-end into the receiver-end of the nodes. Currently, there are advanced hardware designs which can suppress the SI by about 110 dB in certain scenarios, see \cite{7024120}. On the other hand, in the HD mode, transmission and reception take place in the same frequency band but in different time slots, or in the same time slot but in different frequency bands, which avoids the creation of SI. However, since a FD node  uses the resources twice as much as compared to a HD node, the achievable data rates of a network comprised of FD nodes may be significantly higher than that comprised of HD nodes. Motivated by this, in this paper we investigate a network comprised of FD nodes, while, as a special case, we also obtain the optimal centralized D-TDD for a network comprised of HD nodes.

D-TDD schemes  have been investigated in
\cite{6666413,7070655,1705939,4556648,1638665,7136469,4524858,8611365,1261897,8812711,8004461,7000558,7876862}, where \cite{6666413,7070655,1705939,4556648,1638665,7136469,4524858} investigate distributed D-TDD schemes and centralized D-TDD schemes are investigated in \cite{8611365,1261897,8812711,8004461,7000558,7876862}. The works in \cite{8611365,1261897,8812711} propose  non-optimal heuristic centralized scheduling schemes. Specifically, the authors in \cite{8611365} propose a centralized D-TDD scheme named SPARK that provides more than 120$\%$ improvement compared to similar distributed D-TDD schemes.  In \cite{1261897} the authors proposed a centralized D-TDD scheme but do not provide a  mathematical analysis of the proposed scheme. In  \cite{8812711}, the authors applied a centralized D-TDD scheme to optimise the power of the network nodes in order to reduce the inter-cell interference, however, the proposed solution is sub-optimal. The work in  \cite{8004461} proposes a centralized D-TDD scheme for a wireless network where the decisions for transmission and reception at the nodes are chosen from  a finite and predefined set of configurations, which is not optimal in general and may limit the network performance. A network comprised of two-way links  is investigated in \cite{7000558}, where each link can be used either for transmission or reception in a given time slot, with the aim of optimising the direction of the two-way links in each time slot. However, the difficulty of the problem in \cite{7000558} also  leads to a  sub-optimal solution being proposed. The work in \cite{7876862}   investigates a wireless network, where the nodes can select to  transmit, receive, or be silent in a given time slot. However,  the  proposed solution in \cite{7876862}  is again sub-optimal due to the difficulty of the investigated problem. On the other hand, \cite{7801002,7491359} investigate centralized D-TDD schemes for a wireless network comprised of FD nodes. Specifically, the authors in \cite{7801002}
used an approximation to develop a non-optimum game theoretic centralized D-TDD scheme, which uses round-robin scheduling, and they provide analysis for a cellular network comprised of two cells. In  \cite{7491359}, the authors investigate a sub-optimal centralized D-TDD scheme that performs FD and HD mode selection at the nodes based on geometric programming.

To the best of our knowledge, the  optimal centralized D-TDD scheme for a wireless network comprised of FD or HD nodes is an open problem in the literature. As a result, in this paper, we derive the  optimal centralized D-TDD  scheme  for a wireless network comprised of  FD  nodes. In particular, we derive the  optimal  scheduling of the reception, transmission, simultaneous reception and transmission, or silence at every FD node in a given time slot such that the rate region of the network is maximized. In addition, as a special case, we also derive the optimal centralized D-TDD scheme for a  network comprised of   HD  nodes as well as a network comprised of FD and HD nodes. Our numerical results show that the proposed optimal  centralized D-TDD  scheme achieves significant gains over existing  centralized D-TDD schemes.

The rest of this paper is organized as follows. In Section \ref{Sec-Sys}, we present the system and channel model. In Section \ref{PRoblem_def}, we formulate the centralized D-TDD problem. In  Section \ref{Sec-DTDD}, we present the optimal centralized D-TFDD scheme for a wireless network comprised of  FD and HD nodes. In Section  \ref{Sec-QoS}, we investigate rate allocation fairness and propose a corresponding rate allocation scheme. Simulation and numerical results are provided in Section \ref{Sec-Num}, and the conclusions are drawn in Section \ref{Sec-Conc}.


 \begin{figure}[t]
\centering\includegraphics[width=5.8in]{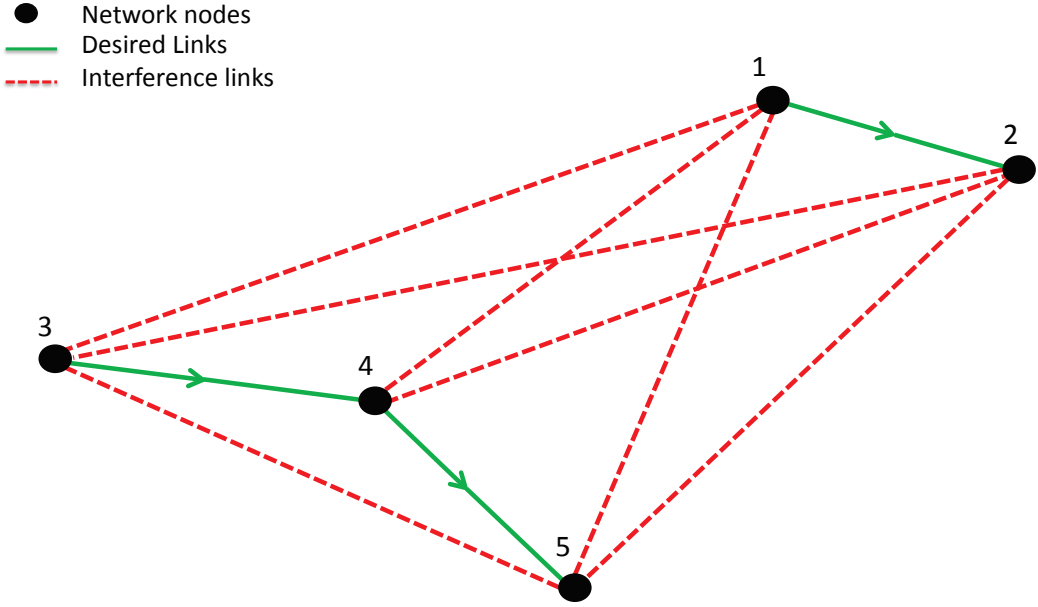}
\vspace{-8mm}
\caption{A wireless network comprised of 5   nodes.}
\label{general_networ_Sys_Model}
\end{figure}

\section{System Model}\label{Sec-Sys}
In this section, we present the system  and channel models.

\subsection {System Model}
We consider a wireless network comprised of $K$ FD nodes. Each network node is be able to wirelessly communicate with the rest of the nodes in the network and in a given time slot operate as: 1) a receiver that receives information from other network nodes, 2) a transmitter that sends information to other network nodes, 3) simultaneously receive and transmit information from/to other network nodes, or 4) be silent.  The nodes can change their state from one time slot to the next.
Moreover, in the considered network, we assume that each  node is able to receive information from multiple  nodes simultaneously utilizing a multiple-access channel scheme, see \cite[Ch.~15.1.2]{cover}, however,  a node cannot transmit  information to more than one node, i.e., we assume that information-theoretic broadcasting   schemes, see \cite[Ch.~15.1.3]{cover}, are not employed. Hence, the considered network is a collection of many multiple-access channels all operating in the same frequency band.

In the considered wireless  network, we assume that there exist a link between any two nodes in the network, i.e., that the network graph is a complete graph. Each link is assumed to be impaired by independent flat fading, which is modelled via the channel gain of the link. The channel gain between any two nodes can be set to  zero during the entire transmission time, which in turn models the case when the wireless signal sent from one of the two nodes can not propagate and reach the other node. Otherwise, if the channel gain is non-zero in any time slot during the transmission, then  the wireless signal sent from one of the two nodes can reach the other node. Obviously, not all of the links leading to a given node carry desired information and are thereby desired by the considered node. There are links which carry undesired information to a considered node, which are referred to as interference links. An interference link causes the signal transmitted from a given node to reach an  unintended destination node, and acts as interference to that node. For example, in Fig.~\ref{general_networ_Sys_Model}, node 2 wants to receive information from node 1. However, since nodes 3 and 4  are also transmitting  in the same time slot, node 2 will experience interference from nodes 3 and 4. Similarly, nodes 4 and 5 experience interference from node 1. It is easy to see that for node 2 it is beneficial if all other nodes, except node 1, are either receiving or silent. However, such a scenario would be  harmful for the rest of the network nodes since they will not be able to receive and transmit any data. 

In order to model the desired and undesired links for each node, we introduce a binary matrix  $\mathbf{ Q}$ defined as follows. The  $(j,k)$ element of $\mathbf{ Q}$ is equal to 1  if node $k$ regards the signal transmitted from node $j$ as a desired signal, and is equal to 0 if  node $k$ regards the signal transmitted from node $j$ as an interference signal.   Moreover, let $\mathbf{\bar Q}$ denote an identical matrix as  $\mathbf{ Q}$ but with flipped binary values. Hence, the  $(j,k)$ element of $\mathbf{\bar Q}$ assumes the value 1 if node $k$ regards the signal transmitted from node $j$ as interference, and the  $(j,k)$ element of $\mathbf{\bar Q}$ is 0 when node $k$ regards the signal transmitted from node $j$ as a desired signal.

The matrix $\mathbf{Q}$, and thereby also the matrix $\mathbf{\bar Q}$, are set before the start of the transmission in the network. How a receiving node decides from which nodes it receives desired signals, and thereby from which node it receives interference signals, is unconstrained for the analyses in this paper.

\subsection{Channel Model}
We assume that each node in the considered network is impaired by unit-variance additive white Gaussian noise (AWGN), and that the links between the nodes are impaired by block fading. In addition,  due to the in-band simultaneous reception-transmission, each node is also impaired by SI, which occurs due to leakage of energy from the transmitter-end into the receiver-end of the node. The SI impairs the decoding of the received information signal significantly, since the SI signal has a relatively higher power compared to the power of the desired signal. Let the transmission on the network be carried-out over $T\to\infty$ time slots, where a time slot is small enough such that  the fading on all network links, including the SI links, can be considered constant during  a time slot. Hence, the instantaneous signal-to-noise-ratios (SNRs) of the links are assumed to change only from one time slot to the next and not within a time slot. Let $g_{j,k}(i)$  denote the fading coefficient of the channel between nodes $j$ and $k$ in the considered network in time slot $i$. Then $\gamma_{j,k}(i)={|g_{j,k}(i)|^2}$ denotes the instantaneous SNR of the channel between nodes $j$ and $k$, in time slot $i$. The case when $j=k$  models the SNR of the SI channel of node $k$ in time slot $i$, given by $\gamma_{j,k}(i)={|g_{j,k}(i)|^2}$. Note that, since the links are impaired by fading, the values of $\gamma_{j,k}(i)$ change from one time slot to the next. All the CSIs, $\gamma_{j,k}(i), \forall i,j$ should be aggregated at the central node. 

Finally, let $\mathbf{G}(i)$ denote the weighted connectivity  matrix of the graph of the considered network in time slot $i$, where the $(j,k)$ element in the matrix  $\mathbf{G}(i)$ is  equal to the instantaneous SNR of the link $(j,k)$, $\gamma_{j,k}(i)$. 

\subsection{Rate Region}
Let $SINR_k(i)$ denote the  signal-to-interference-plus-noise-ratio (SINR) at   node $k$ in time slot $i$. Then,  the average rate received at node $k$ over $T\to\infty$ time slots  is given by
\begin{align}\label{eq_11aint}
\bar R_k=\lim_{T\to\infty}\frac{1}{T}\sum_{i=1}^T  \log_2 \left(1+SINR_k(i) \right).
\end{align}
Using  (\ref{eq_11aint}), $\forall k$, we define a weighted sum-rate as
\begin{align}\label{eq_11bint}
\Lambda= \sum_{k=1}^K  \mu_k \bar R_k,
\end{align}
where the values of $\mu_k$, for $0 \leq \mu_k \leq 1$, $\sum_{k=1}^K  \mu_k=1$ are fixed. By maximizing (\ref{eq_11bint}) for any fixed $\mu_k$,  $\forall k$, we obtain one point of the boundary of the rate region. All possible values of $\mu_k, \forall k$, provide all possible values of the boundary line of the rate region of the network.

\section{Problem Formulation}\label{PRoblem_def}
Each node in the network can be in one of the following four states: receive ($r$), transmit ($t$), simultaneously receive and transmit ($f$), and silent ($s$). The main problem in the considered wireless network is to find the optimal state of each node in the network in each time slot, based on global knowledge of the channel fading gains, such that the  weighted sum-rate of the network,  given by (\ref{eq_11bint}), is maximized. To model the modes of each node in each time slot, we define the following binary variables for node $k$ in time slot $i$
\begin{align}
r_k(i)&=\left\{
\begin{array}{ll}\label{eq_1}
1 &\textrm{if node } k \textrm{ receives in time slot } i\\
0 & \textrm{otherwise},
\end{array}
\right.\\
t_k(i)&=\left\{
\begin{array}{ll}\label{eq_2}
1 &\textrm{if node } k \textrm{ transmits in time slot } i\\
0 & \textrm{otherwise},
\end{array}
\right.\\
f_k(i)&=\left\{
\begin{array}{ll}\label{eq_3}
1 &\textrm{if node } k \textrm{ simultaneously receives and transmits  in time slot } i\\
0 & \textrm{otherwise},
\end{array}
\right.\\
s_k(i)&=\left\{
\begin{array}{ll}\label{eq_41}
1 &\textrm{if node } k \textrm{ is silent in time slot } i\\
0 & \textrm{otherwise}.
\end{array}
\right.
\end{align}
Since node $k$ can be in one and only one mode in each time slot, i.e., it can either receive, transmit, simultaneously receive and transmit, or be silent, the following has to hold
\begin{align}\label{eq_4}
r_k(i)+t_k(i)+f_k(i)+s_k(i)=1,\;\forall k.
\end{align}
For the purpose of simplifying the analytical derivations, it is more convenient to represent (\ref{eq_4}) as
\begin{align}\label{eq_5}
r_k(i)+t_k(i)+f_k(i)\in\{0,1\},\;\forall k,
\end{align}
where if $r_k(i)+t_k(i)+f_k(i)=0$ holds, then node $k$ is silent in time slot $i$.

Now, using the binary variables defined in (\ref{eq_1})-(\ref{eq_41}), we define vectors  $\mathbf{r}(i)$, $\mathbf{t}(i)$, $\mathbf{f}(i)$, and $\mathbf{s}(i)$ as
\begin{align}
\mathbf{r}(i)&=[r_1(i),\; r_2(i),\;...,\; r_K(i)],\label{eq_7a}\\
\mathbf{t}(i)&=[t_1(i),\; t_2(i),\;...,\; t_K(i)],\label{eq_7b}\\
\mathbf{f}(i)&=[f_1(i),\; f_2(i),\;...,\; f_K(i)],\label{eq_7c}\\
\mathbf{s}(i)&=[s_1(i),\; s_2(i),\;...,\; s_K(i)].\label{eq_7d}
\end{align}
Hence, the $k$-th element of the vector $\mathbf{r}(i)$/$\mathbf{t}(i)$/$\mathbf{f}(i)$/$\mathbf{s}(i)$  is $r_k(i)$/$t_k(i)$/$f_k(i)$/$s_k(i)$, and this element shows whether the $k$-th node is receiving/transmitting/simultaneously receiving and transmitting/silent. Therefore, the four vectors $\mathbf{r}(i)$, $\mathbf{t}(i)$, $\mathbf{f}(i)$, and $\mathbf{s}(i)$, given by (\ref{eq_7a})-(\ref{eq_7d}),  show which nodes in the network are  receiving, transmitting, simultaneously receiving and transmitting, and are silent in time slot $i$, respectively. Due to condition (\ref{eq_4}), the elements in the   vectors $\mathbf{r}(i)$, $\mathbf{t}(i)$, $\mathbf{f}(i)$, and $\mathbf{s}(i)$ are mutually dependent and have to satisfy the following condition 
\begin{equation}
\mathbf{r}(i)+\mathbf{t}(i)+\mathbf{f}(i)+\mathbf{s}(i)=\mathbf{e},
\end{equation}
where $\mathbf{e}$ is the all-ones vector, i.e., $\mathbf{e}=[1,1,...,1]$.

The main problem in the considered wireless network  is finding
the optimum vectors $\mathbf{r}(i)$, $\mathbf{t}(i)$, $\mathbf{f}(i)$, and $\mathbf{s}(i)$  that maximize  the boundary of the rate region of the network,  which can be obtained  by using the following optimization problem
\begin{align}
& {\underset{\mathbf{r}(i), \mathbf{t}(i), \mathbf{f}(i), \mathbf{s}(i),\forall i} {  \textrm{Maximize:} }}\; 
 \Lambda \nonumber\\
&{\rm{Subject\;\;  to \; :}}  \nonumber\\
&\qquad\qquad{\rm C1:}\;  t_v(i)\in\{0,1\}, \; \forall v\nonumber\\
&\qquad\qquad{\rm C2:}\;  r_v(i)\in\{0,1\}, \; \forall v\nonumber\\
&\qquad\qquad{\rm C3:}\;  f_v(i)\in\{0,1\}, \; \forall v\nonumber\\
&\qquad\qquad{\rm C4:}\;  s_v(i)\in\{0,1\}, \; \forall v\nonumber\\
&\qquad\qquad{\rm C5:}\;  s_v(i)+r_v(i)+f_v(i)+s_v(i)=1, \; \forall v,
\label{eq_max_1ab}
\end{align}
where $ \mu_k$ are fixed. The solution of this problem is given in   Theorem~\ref{theo_2} in Section~\ref{Sec-DTDD}.

Before investigating the problem in (\ref{eq_max_1ab}),  we define two auxiliary matrices that will help us derive the main result. Specifically, using matrices $\mathbf{G}(i)$,  $\mathbf{Q}$, and $\mathbf{\bar Q}$ defined in Sec.~\ref{Sec-Sys}, we define two auxiliary matrices $\mathbf{D}(i)$ and $\mathbf{I}(i)$, as
\begin{align}
\mathbf{D}(i) &= \mathbf{G}(i)\circ \mathbf{Q},\label{eq_10a}\\
\mathbf{I}(i) &= \mathbf{G}(i)\circ \mathbf{\bar Q}\label{eq_10b},
\end{align}
where $\circ$ denotes the Hadamard product of matrices, i.e., the element wise multiplication of two matrices. Hence, elements in the matrix $\mathbf{D}(i)$ are the instantaneous SNRs of the desired links which carry desired information. Conversely, the elements in the matrix $\mathbf{I}(i)$ are the instantaneous SNRs  of the interference links which carry undesired information. Let $\mathbf{d}_k^{\intercal}(i)$ and $\mathbf{i}_k^{\intercal}(i)$ denote the $k$-th column vectors of the matrices $\mathbf{D}(i) $ and $\mathbf{I}(i)$, respectively. The vectors $\mathbf{d}_k^{\intercal}(i)$ and $\mathbf{i}_k^{\intercal}(i)$  show the instantaneous SNRs  of the desired and interference links for node $k$ in time slot $i$, respectively. For example, if the third and fourth elements in $\mathbf{d}_k^{\intercal}(i)$ are non-zero and thereby equal to $\gamma_{3,k}(i)$ and $\gamma_{4,k}(i)$, respectively, then this means that the $k$-th node receives desired signals from the third and the fourth elements in the network via channels which have squared instantaneous SNRs  $\gamma_{3,k}(i)$ and $\gamma_{4,k}(i)$, respectively. Similar, if the fifth, sixth, and $k$-th   elements in $\mathbf{i}_k^{\intercal}(i)$ are non-zeros and thereby equal to $\gamma_{5,k}(i)$, $\gamma_{6,k}$ and $\gamma_{k,k}(i)$, respectively, it means that the $k$-th node receives interference  signals from the fifth and the sixth nodes in the network via channels which have squared instantaneous SNRs  $\gamma_{5,k}(i)$ and $\gamma_{6,k}(i)$, respectively, and that  the $k$-th node suffers from SI with squared instantaneous SNR $\gamma_{k,k}(i)$.

\begin{remark}
A central processor is assumed to collect all instantaneous SNRs, $\gamma_{j,k}(i)$, and thereby construct $\mathbf{G}(i)$ at the start of time slot $i$. This central unit will then decide the optimal values of $\mathbf{r}(i)$, $\mathbf{t}(i)$, $\mathbf{f}(i)$ and $\mathbf{s}(i)$, defined in (\ref{eq_7a})-(\ref{eq_7d}), based on the proposed centralized D-TDD scheme, and broadcast these values to the rest of the nodes. Once the optimal values of $\mathbf{r}(i)$, $\mathbf{t}(i)$, $\mathbf{f}(i)$, and $\mathbf{s}(i)$ are known at all nodes the transmissions,  receptions, simultaneous transmission and reception, and silences of the nodes can start in time slot $i$. Obviously, acquiring  global CSI at a central processor is impossible in practice as it will incure a huge overhead and, by the time it is used, the CSI will likely be outdated. However,  this assumption will allow us to compute an upper bound on the network performance which will serve as an upper bound to the performance of any  D-TDD scheme.
\end{remark}

\begin{remark}
Note that the optimal state of the nodes of the network (i.e., receive, transmit, simultaneously receive and transmit, or silent) in each time slot can also be obtained by brute-force search. Even if this is possible for a small network, an analytical solution of the problem will provide depth insights into the corresponding problem.
\end{remark}

\remark{
In this paper, we only optimize the reception-transmission schedule of the nodes, and not the transmission coefficients  of the nodes, which leads to  interference alignment \cite{4533226}. Combining adaptive reception-transmission with interference alignment is left for future work.
}

\section{The Optimal Centralized D-TDD Scheme}\label{Sec-DTDD}
Using the notations in Sections\ref{Sec-Sys} and~\ref{PRoblem_def}, we state a theorem that models the received rate at node $k$ in time slot $i$.
\begin{theorem}\label{theo_1}
Assuming that all nodes transmit with power $P$,  then the received rate at node $k$ in time slot $i$ is given by
\begin{align}\label{Eqe_SINR_3as}
R_k(i)=\log_2\left(1+ [r_k(i)+f_k(i)] \frac{P\, \left [ \mathbf{t}(i)+\mathbf{f}(i) \right]  \mathbf{d}_k^{\intercal}(i)}{1+ P\, \left [\mathbf{t}(i)+\mathbf{f}(i) \right]  \mathbf{i}_k^{\intercal}(i)}\right),
\end{align} 
which is achieved by a multiple-access channels scheme between the desired nodes of node $k$ acting as transmitter and node $k$ acting as a receiver. To this end,
 node $k$ employs successive interference cancellation to the codewords from the desired nodes whose rates are appropriately adjusted in order for (\ref{Eqe_SINR_3as}) to hold.
\end{theorem}
\begin{IEEEproof}
Please refer to Appendix~\ref{app_PR_Local} for the proof.
\end{IEEEproof}

In (\ref{Eqe_SINR_3as}), we have obtained a very simple and compact expression for the received rate at each  node of the network in each time slot. As can be seen from (\ref{Eqe_SINR_3as}), the rate  depends on the fading channel gains of the desired links via $\mathbf{d}_k^{\intercal}(i)$ and the interference links via $\mathbf{i}_k^{\intercal}(i)$, as well as the state selection vectors of the network via $\mathbf{t}(i)$, $\mathbf{r}(i)$, and $\mathbf{f}(i)$. 

Using the received rate at each  node of the network, defined by (\ref{Eqe_SINR_3as}), we obtain the average received rate at node $k$ as
\begin{align}\label{eq_11a}
\bar R_k=\lim_{T\to\infty}\frac{1}{T}\sum_{i=1}^T  R_k(i), \forall k.
\end{align}

 Inserting (\ref{Eqe_SINR_3as}) into (\ref{eq_11a}), and then (\ref{eq_11a}) into (\ref{eq_11bint}), we obtain the weighted  sum-rate  of the network as 

\begin{align}\label{eq_11c}
\Lambda= \lim_{T\to\infty}\frac{1}{T} \sum_{i=1}^T \sum_{k=1}^N     \mu_k \log_2 \left(1+\left [r_k(t)+f_k(t) \right]\frac{P\, \left [ \mathbf{t}(i)+\mathbf{f}(i) \right]  \mathbf{d}_k^{\intercal}(i)}{1+ P\, \left [\mathbf{t}(i)+\mathbf{f}(i) \right]  \mathbf{i}_k^{\intercal}(i)} \right).
\end{align}
Now, note that the only variables that can be manipulated in  (\ref{eq_11c}) in each time slot are the values of the elements in the vectors $\mathbf{t}(i)$, $\mathbf{r}(i)$, and $\mathbf{f}(i)$, and the values of $\mu_k, \forall k$.  We use $\mathbf{t}(i)$, $\mathbf{r}(i)$, and $\mathbf{f}(i)$ to maximize  the boundary of the rate region for a given  $\mu_k, \forall k$, in the following. In addition, later on in Section \ref{Sec-QoS}, we use the constants $\mu_k, \forall k$, to establish a scheme that achieves fairness between the nodes of the network.

The optimum vectors $\mathbf{r}(i)$, $\mathbf{t}(i)$, $\mathbf{f}(i)$, and $\mathbf{s}(i)$  that maximize  the boundary of the rate region of the network  can be obtained  by the following optimization problem
\begin{align}
& {\underset{\mathbf{r}(i), \mathbf{t}(i), \mathbf{f}(i), \mathbf{s}(i),\forall i} {  \textrm{Maximize:} }}\; 
 \lim_{T\to\infty}\frac{1}{T} \sum_{i=1}^T \sum_{k=1}^N     \mu_k \log_2 \left(1+\left [r_k(t)+f_k(t) \right]\frac{P\, \left [ \mathbf{t}(i)+\mathbf{f}(i) \right]  \mathbf{d}_k^{\intercal}(i)}{1+ P\, \left [\mathbf{t}(i)+\mathbf{f}(i) \right]  \mathbf{i}_k^{\intercal}(i)} \right) \nonumber\\
&{\rm{Subject\;\;  to \; :}}  \nonumber\\
&\qquad\qquad{\rm C1:}\;  t_v(i)\in\{0,1\}, \; \forall v\nonumber\\
&\qquad\qquad{\rm C2:}\;  r_v(i)\in\{0,1\}, \; \forall v\nonumber\\
&\qquad\qquad{\rm C3:}\;  f_v(i)\in\{0,1\}, \; \forall v\nonumber\\
&\qquad\qquad{\rm C4:}\;  s_v(i)\in\{0,1\}, \; \forall v\nonumber\\
&\qquad\qquad{\rm C5:}\;  s_v(i)+r_v(i)+f_v(i)+s_v(i)=1, \; \forall v,
\label{eq_max_1a}
\end{align}
where $ \mu_k$ are fixed. The solution of this problem is given in the following theorem.
\begin{theorem}\label{theo_2}
 The optimal values of the vectors $\mathbf{t}(i)$, $\mathbf{r}(i)$, $\mathbf{f}(i)$, and $\mathbf{s}(i)$, which maximize  the boundary of the rate region of the network, found as  the solution of (\ref{eq_max_1a}), is given by Algorithm~\ref{Lyp_algorithm}, which is explained in details in the following.

\begin{algorithm}
\caption{Finding the optimal vector, $\mathbf{t}(i)$}
\label{Lyp_algorithm}
\begin{algorithmic}[1]
\Procedure{ $\forall \;i \in \{1,2,...,T\}$}{}\label{lyp_proc}
\State Initiate  $n=0$, and $t_x(i)_0$,  $w_x(i)_0$ and $l_x(i)_0$ randomly, $x \in \{1,2,..,K\}$, where $\mathcal{K}_d$  is the set of desired nodes.
\State \textrm{****** Iterative-loop starts*****}
\While{exit-loop-flag $==$ FALSE}
\If{ (\ref{eq_outer_T19}) holds} \label{cvx_found}
\State exit-loop-flag $\gets$ TRUE
\Else
\State \label{Lable_1} n++
\State compute $t_x(i)_n$  with (\ref{eq_scheme_D-TDD})
\State compute $w_x(i)_n$  with (\ref{eq_max_app_T4})
\State compute $l_x(i)_n$ with (\ref{eq_max_app_T7})
\EndIf
\EndWhile
\State \textrm{****** Iterative-loop end*****}
\If{ $t_x(i)=0$ and $t_k(i)=1, \forall k$,  where $(x,k)$ element of $\mathbf{ Q}$ is one }
\State $r_x(i)=1$
\EndIf
\If{ $t_x(i)=0$ and $t_k(i)=0, \forall k$,  where $(x,k)$ element of $\mathbf{ Q}$ is one }
\State $s_x(i)=1$
\EndIf
\If{ $t_x(i)=1$ and $t_k(i)=1, \forall k$,  where $(x,k)$ element of $\mathbf{ Q}$ is one }
\State $f_x(i)=1$ and $t_x(i)=0$
\EndIf
\If{ $t_x(i)=1$ and $t_k(i)=0, \forall k$,  where $(x,k)$ element of $\mathbf{ Q}$ is one }
\State $t_x(i)$ remains unchanged
\EndIf

\State \Return $\mathbf{t}(i)$, $\mathbf{r}(i)$, $\mathbf{f}(i)$, $\mathbf{s}(i)$  
\EndProcedure
\end{algorithmic}
\end{algorithm}


Algorithm~\ref{Lyp_algorithm}  is an iterative algorithm. Each iteration has its own index, denoted by $n$. In each iteration, we compute the vector $\mathbf{t}(i)$ in addition to two auxiliary vectors $\mathbf{w}(i)=\{w_1(i),w_2(i),...$ $,w_N(i)\}$ and $\mathbf{l}(i)=\{l_1(i),l_2(i),...,l_N(i)\}$. Since the   computation process is iterative, we add the index $n$ to denote the $n$'th iteration. Hence, the  variables $t_x(i)$, $w_x(i)$, and $l_x(i)$ in iteration $n$ are denoted by  $t_x(i)_n$,  $w_x(i)_n$, and $l_x(i)_n$, respectively. In each iteration, $n$, the variable $t_x(i)_{n}$, for $x \in \{1,2,...,N\}$, is  calculated as
\begin{align}\label{eq_scheme_D-TDD}
\bullet \; t_x(i)_{n}=0 & \; \textrm{  if}\;\; \sum_{k=1}^N  \frac{P \mu_k l_k(i)_{n-1}}{\ln{2}} \hspace{-1.5mm} \left[  d_{x,k}(i) \hspace{-1.0mm} \left (\hspace{-1.0mm}1\hspace{-1.0mm}-\hspace{-1.0mm}\frac{w_{k}(i)_{n-1}}{\sqrt{P\sum_{v=1,v \ne x}^N {t_v(i)_{n-1} d_{v,k}(i)} }} \right)+i_{x,k}(i)  \right] \geq 0, \nonumber\\
\bullet \; t_x(i)_{n}=1 & \;\textrm{  if}\;\; \textrm{otherwise}.  
\end{align}
In (\ref{eq_scheme_D-TDD}),  $d_{v,k}(i)$ and $i_{v,k}(i)$ are the $(v,k)$ elements of the matrices $\mathbf{D}(i)$ and $\mathbf{I}(i)$, respectively. Whereas,  $l_k(i)_n$ and $w_{k}(i)_n$ are the  auxiliary  variables, and they are treated as constants in this stage and will be given  in the following.

In iteration $n$, the variable   $w_{x}(i)_n$, for $x \in \{1,2,...,N\}$, is  calculated  as
 \begin{align}
w_{x}(i)_n=&\frac{{A}_{x}(i)+{B}_{x}(i)}{\sqrt{{A}_{x}(i)}},
\label{eq_max_app_T4}
\end{align} 
where ${A}_{x}(i)$ and ${B}_{x}(i)$ are defined as
\begin{align}
{A}_{x}(i)&= P  \mathbf{t}(i)_n  \mathbf{d}_x^{\intercal}(i),\label{eq_T7a}\\
{B}_{x}(i)&= 1+P  \mathbf{t}(i)_n  \mathbf{i}_x^{\intercal}(i).\label{eq_T7b}
\end{align}

In iteration $n$, the variable  $l_x(i)_n$, for $x \in \{1,2,...,N\}$, is  calculated  as
 \begin{align}
l_x(i)_n=&\frac{1}{\left( |\sqrt{{A}_{x}(i)}-w_{x}(i)_n|^2+{B}_{x}(i)  \right)},
\label{eq_max_app_T7}
\end{align} 
where  $w_{x}(i)_n$ is treated as constant in this stage. In addition, ${A}_{x}(i)$ and ${B}_{x}(i)$, are given by (\ref{eq_T7a}) and (\ref{eq_T7b}), respectively.

The process of updating the variables $t_x(i)_n$,  $w_x(i)_n$, and $l_x(i)_n$ for each time slot $i$ is repeated until convergence occurs, which can be checked by the following equation
\begin{align}\label{eq_outer_T19}
|\Lambda_n - \Lambda_{n-1}| < \epsilon,
\end{align}
where $\Lambda_n=\sum_{k=1}^N \mu_k \log_2 \left(1+\frac{P  \mathbf{t}(i)_n  \mathbf{d}_k^{\intercal}(i)}{1+P  \mathbf{t}(i)_n  \mathbf{i}_k^{\intercal}(i)} \right)$. Moreover, $\epsilon>0$ is a relatively small constant, such as $\epsilon=10^{-6}$.

Once $t_x(i), \forall x$, is decided,  the other variables, $r_x(i)$, $f_x(i)$, and $s_x(i)$ can be calculated as follows. If  $t_x(i)=0$, $t_k(i)=1$, and the $(x,k)$ element of $\mathbf{ Q}$ is equal to one, then  $r_x(i)=1$.  If  $t_x(i)=0$, $t_k(i)=0$, and $(x,k)$ element of $\mathbf{ Q}$ is equal to one, then  $s_x(i)=1$.   If  $t_x(i)=1$, $t_k(i)=1$, and $(x,k)$ element of $\mathbf{ Q}$ is equal to one, then  $f_x(i)=1$ and we set $t_x(i)=0$. Finally, if $t_x(i)=1$, $t_k(i)=0$, and $(x,k)$ element of $\mathbf{ Q}$ is equal to one, then  $t_x(i)$ remains unchanged.

\end{theorem}
\begin{IEEEproof}
Please refer to Appendix~\ref{app_PR_Outage_L} for the proof.
\end{IEEEproof}

\subsection{Special Case of the Proposed Centralized D-TDD Scheme for HD nodes}

As a special case of the proposed centralized D-TDD scheme for a network comprised of FD nodes proposed in Theorem~\ref{theo_2}, we investigate the optimal centralized D-TDD scheme for network comprised of HD nodes that maximizes the rate region.

For the case of a network comprised of HD nodes, we again use the vectors $\mathbf{r}(i)$, $\mathbf{t}(i)$, and $\mathbf{s}(i)$, and set the vector $\mathbf{f}(i)$ to all zeros due to the HD mode. The optimum vectors $\mathbf{r}(i)$, $\mathbf{t}(i)$, and $\mathbf{s}(i)$  that maximize the boundary of the rate region of a network comprised of HD nodes  can be obtained  by the following optimization problem
\begin{align}
& {\underset{\mathbf{r}(i), \mathbf{t}(i), \mathbf{s}(i),\forall i} {  \textrm{Maximize:} }}\; 
 \lim_{T\to\infty}\frac{1}{T} \sum_{i=1}^T \sum_{k=1}^N     \mu_k \log_2 \left(1+r_k(t)\frac{P\,  \mathbf{t}(i)  \mathbf{d}_k^{\intercal}(i)}{1+ P\, \mathbf{t}(i)  \mathbf{i}_k^{\intercal}(i)} \right) \nonumber\\
&{\rm{Subject\;\;  to \; :}}  \nonumber\\
&\qquad\qquad{\rm C1:}\;  t_v(i)\in\{0,1\}, \; \forall v\nonumber\\
&\qquad\qquad{\rm C2:}\;  r_v(i)\in\{0,1\}, \; \forall v\nonumber\\
&\qquad\qquad{\rm C3:}\;  s_v(i)\in\{0,1\}, \; \forall v\nonumber\\
&\qquad\qquad{\rm C4:}\;  s_v(i)+r_v(i)+s_v(i)=1, \; \forall v,
\label{eq_max_1a3}
\end{align}
where $\mu_k, \forall k$ is fixed. The solution of this problem is given in the following theorem.
\begin{theorem}\label{theo_3}
The optimal values of the vectors $\mathbf{t}(i)$, $\mathbf{r}(i)$, and $\mathbf{s}(i)$ which maximize  the boundary of the  rate region of the considered  network comprised  of HD nodes, found as  the solution of (\ref{eq_max_1a3}), is also given by Algorithm~\ref{Lyp_algorithm} where
 lines 17-18 in  Algorithm~\ref{Lyp_algorithm} need to be removed and where $\gamma_{j,k}(i)$ is set to  $\gamma_{j,k}(i)=\infty, \forall j=k$ and $\forall i$ in the weighted connectivity  matrix $\mathbf{G}(i)$.

\end{theorem}
\begin{IEEEproof}
Please refer to Appendix~\ref{app_PR_Outage_L3} for the proof.
\end{IEEEproof}

\begin{remark}
For the case when the network is comprised of both FD and HD nodes, Theorem~\ref{theo_3} needs to be applied only to the HD nodes in order to obtain the optimal centralized D-TDD scheme for this case
\end{remark}

\section{Rate Allocation Fairness}\label{Sec-QoS}
The nodes in a network have different rate demands based on the application they employ. In this section, we propose a scheme that allocates resources to the network nodes based on the rate demand of the network nodes. To this end, in the following, we assume that the central processor has access to the  rate demands of the network nodes.

Rate allocation can be done using a  prioritized rate allocation policy, where some nodes have a higher priority compared to others, and thereby, should be served preferentially. For example, some nodes are paying more  to the network operator compared to the other nodes in exchange for higher data rates. In this policy, nodes with lower priority are served only when higher priority nodes are served acceptably. On the other hand, nodes that have the same priority level should be served by a fair rate allocation scheme that allocates resources proportional to the node needs.

In the optimal centralized D-TDD scheme given in Theorem~\ref{theo_2}, the average received rate of  user $k$ can be controlled via the constant $0\leq \mu_k\leq 1, \forall k $.  By varying $\mu_k$ from zero to one, the average received rate of user $k$ can be increased from zero to the maximum possible rate. Thereby, by optimizing the value of $\mu_k, \forall k$, we can establish a rate allocation scheme among the users which allocates resources  based on the rate demand of the nodes. In the following, we propose a practical centralized D-TDD scheme for  rate allocation in real-time by adjusting the values of  $\boldsymbol{\mu}=[\mu_1,\; \mu_2,\;...,\; \mu_K]$.

\subsection{Proposed Rate Allocation Scheme For a Given Fairness}\label{Sec-Fair-pri}

The average received rate at  node $k$ obtained using the proposed optimal centralized D-TDD scheme is given by
\begin{align}\label{eq_max_Fairness_12a}
\bar R_k(\boldsymbol{\mu})=\lim_{T\to\infty} \frac{1}{T}\sum_{i=1}^T R_k^*(i,\boldsymbol{\mu}),
\end{align}
where $R_k^*(i,\mu_k)$  is the  maximum received  rate at node $k$ in the time slot $i$, obtained by Algorithm~\ref{Lyp_algorithm} for fixed $\boldsymbol{\mu}$. 

Let $\boldsymbol{\tau} =[\tau_1,\; \tau_2,\;...,\; \tau_K]$, where $\tau_k \geq 0$ be a vector of the  rate demands of the nodes and let $\boldsymbol{\alpha}=[\alpha_1,\; \alpha_2,\;...,\; \alpha_K]$ be the priority level vector of the nodes, where  $0\leq \alpha_k\leq 1$ and $\sum_{k=1}^N \alpha_k=1$. The  priority level vector, $\alpha_k$,  determines the importance of user $k$ such that the higher the value of $\alpha_k$, the higher the priority of the $k$-th node.

In order to achieve rate allocations according to the rate demands in  $\boldsymbol{\tau}$ and the priority levels in $\boldsymbol{\alpha}$, we aim to minimize the weighted squared difference between the average received rate $\bar R_k(\boldsymbol{\mu})$ and the  rate demand, given by $\tau_k, \forall k$, i.e., to make the weighted sum squared error, $\sum_{k=1}^N \alpha_k \left ( \bar R_k(\boldsymbol{\mu}) -\tau_k \right )^2$, as smallest as possible. Note that there  may not be enough network resource to make the weighted sum squared error to be equal to zero. However, the higher $\alpha_k$ is, more network resources need to be allocated to node $k$ in order to increase its rate and bring $\bar R_k(\boldsymbol{\mu})$ close to $\tau_k$.

Using $\boldsymbol{\tau}$ and $\boldsymbol{\alpha}$, we   devise the following rate-allocation problem 
\begin{align}
& {\underset{\boldsymbol{\mu}} {  \textrm{Minimize:} }}\; 
 \sum_{k=1}^N \alpha_k\left ( \bar R_k(\boldsymbol{\mu}) -\tau_k \right )^2 \nonumber\\
&{\rm{Subject\;\;  to \; :}}  \nonumber\\
&\qquad\qquad{\rm C1:}\;  0\leq \mu_k\leq 1, \; \forall k \nonumber\\
&\qquad\qquad{\rm C1:}\;  \sum_{k=1}^N \mu_k=1. 
\label{eq_max_Fairness_1a}
\end{align}
The optimization problem in  (\ref{eq_max_Fairness_1a}) belongs to a family of a well investigated optimization problems in \cite{doi:10.1137/1.9781611970920}, which  do not have closed form solutions. Hence, we propose the following heuristic solution  of (\ref{eq_max_Fairness_1a}) by setting $\boldsymbol{\mu}$  to $\boldsymbol{\mu}=\boldsymbol{\mu}^e(i)$, where each element of $\boldsymbol{\mu}^e(i)$ is obtained as
\begin{align}\label{eq_F1_mu_priori}
&\mu_k^e(i+1)=\mu_k^e(i) +\delta_k(i) \alpha_k \left[ \bar R_k^e(i,\boldsymbol{\mu}^e(i)) - \tau_k \right],
\end{align}
where $\delta_k( i)$, $\forall k$, can be some properly chosen monotonically decaying function of $i$ with $\delta_k( 1)<1$, such as $\frac{1}{2i}$. Note that after updating $\mu_k^e(i), \forall k$, values, we should normalize them to bring  $\mu_k^e(i), \forall k$ in the range ($0\leq \mu_k\leq 1$). To this end, we apply the following normalization method

\begin{align}\label{eq_F1_norm}
&\mu_k^e(i+1)= \frac{\mu_k^e(i+1)}{\sum_{k=1}^N \mu_k^e(i+1)}, \forall k. 
\end{align}

In (\ref{eq_F1_mu_priori}), $ \bar R_k^e(i,\boldsymbol{\mu}^e(i))$ is the real time estimation of $\bar R_k(\boldsymbol{\mu})$, which is given by
\begin{align}\label{eq_F1_R1}
\bar R_k^e(i,\boldsymbol{\mu}^e(i))&=\frac{i-1}{i} \bar R_k^e(i-1,\mu_k^e(i-1)) +\frac{1}{i}   R_k^*(i,\boldsymbol{\mu}^e(i)).
\end{align}

\section{Simulation and Numerical Results}\label{Sec-Num}
In this section, we provide numerical results where we compare the proposed optimal centralized D-TDD scheme  with benchmark centralized D-TDD schemes found in the literature.  All of the presented results in this section are generated for Rayleigh fading by numerical evaluation of the derived results and are confirmed by Monte Carlo simulations.

\textit{The Network:} In all numerical examples, we use  a network covering an area of $\rho \times \rho$ $m^2$. In this area, we place 50 pairs of nodes randomly as follows. We randomly place one node of each pair in the considered area and then the paired node is placed by choosing an angle uniformly at random from $0^\circ$ to $360^\circ$ and choosing a distance uniformly at random from $\chi$=10 m to 100 m,  from the first node. For a given pair of two nodes, we assume that only the link between the paired nodes is desired  and all other links  act as interference links. The channel gain corresponding to the each link is assumed to have Rayleigh  fading, where the mean of $\gamma_{j,k}(i)$  is calculated using the standard path-loss model \cite{6847175} as
\begin{eqnarray}
E\{\gamma_{j,k}(i)\} = \left(\frac{c}{{4\pi {f_c}}}\right)^2\chi_{jk}^{ - \beta }\ \textrm{, for } k\in\{U,D\},
\end{eqnarray}
where $c$ is the speed of light, $f_c=1.9$ GHz is the carrier frequency, $\chi_{jk}$ is the distance between node $j$ and $k$, and $\beta=3.6$ is the path loss exponent. In addition, the average SI suppression varies from 110 dB to 130 dB.

\textit{Benchmark Scheme 1 (Conventional scheme):} This benchmark is the TDD scheme used in  current wireless networks. The network nodes are divided into two groups, denoted by A and B. In odd time slots, nodes in group A send information  to the desired nodes in group B. Then, in the even time slots, nodes in group B send information  to the desired nodes in group A. With this approach there is no interference between the nodes within  group A and within group B since the transmissions are synchronized. However, there are interferences from the nodes in group A to the nodes in group B, and vice versa.
  
\textit{Benchmark Scheme 2 (Interference spins scheme):} The interference spins scheme, proposed in \cite{7000558}, has been considered as the second benchmark scheme.

\textit{Benchmark Scheme 3 (Conventional FD scheme):} This benchmark is the TDD scheme used in  a wireless networks with FD nodes. The network nodes are divided into two groups, denoted by A and B. In all the time slots, nodes in group A send information  to the desired nodes in group B, and also nodes in group B send information  to the desired nodes in group A. The SI  suppression is set to 110 dB.

\subsection{Numerical Results}

\begin{figure}[t]
\vspace*{-2mm}
\centering\includegraphics[width=6.7in]{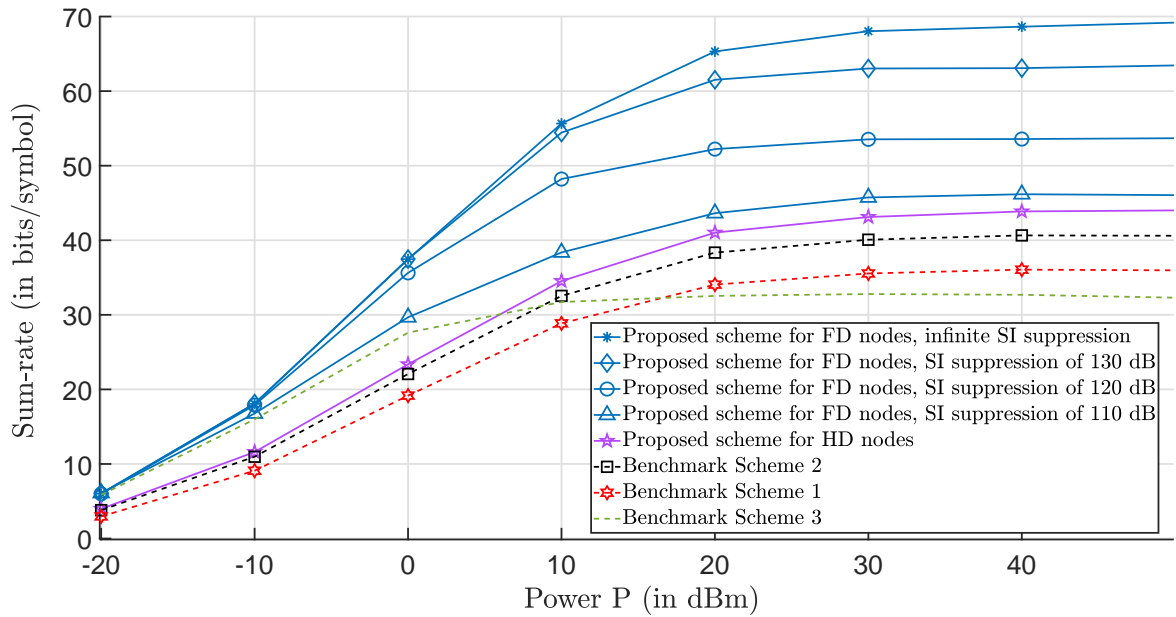}
\caption{Sum-rate vs. transmit power P of the proposed schemes and the benchmark schemes for $\rho$=1000 $m$.}
\label{Rate_Power}
\end{figure}

In Fig.~\ref{Rate_Power}, we show the sum-rates achieved using the proposed scheme for different SI suppression levels and the benchmark schemes as a function of the transmission power at the nodes, $P$. This example is for an area of 1000*1000  $m^2$, where $\mu_k$ is fixed to $\mu_k=\frac{1}{k}, \forall k$. As can be seen from Fig.~\ref{Rate_Power}, for the low transmit power region, where noise is dominant, all schemes  achieve a similar sum-rate. However, increasing the transmit power causes  the overall interference to increase, in which case the optimal centralized D-TDD  scheme achieves a large gain over the considered benchmark schemes. The benchmark schemes show limited performance since in the high power region they can not avoid the interference as effective as the proposed scheme.

\begin{figure}[t]
\vspace*{-2mm}
\centering\includegraphics[width=6.7in]{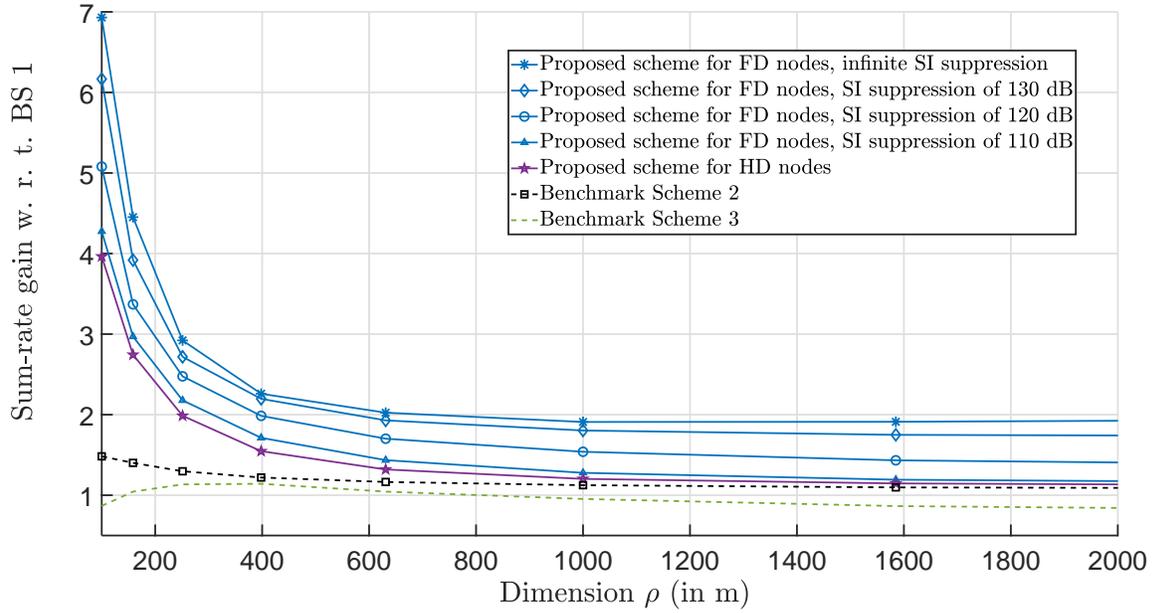}
\caption{Sum-rate vs. dimension D of the proposed schemes and the benchmark schemes for P=20 dBm.}
\label{Rate_Dim}
\end{figure}

 In Fig.~\ref{Rate_Dim}, the sum-rates gain with respect to (w. r. t.) Benchmark Scheme 1 (BS 1) is presented  for different schemes as a function of the dimension of the considered area, $\rho$. We assume that the transmit power is fixed to  $P$=20 dBm, and $\mu_k=\frac{1}{k}, \forall k$. Since the nodes are placed randomly in an area of $\rho \times \rho$ $m^2$, for large $\rho$, the  links become more separated  and the interference has a weeker effect. As a result,  all of the schemes have close sum-rate results. However, decreasing the  dimension, $\rho$, causes  the overall interference to increase, which leads to the optimal centralized D-TDD scheme to have a considerable gain over the benchmark schemes.

\begin{figure}[t]
\vspace*{-2mm}
\centering\includegraphics[width=6.7in]{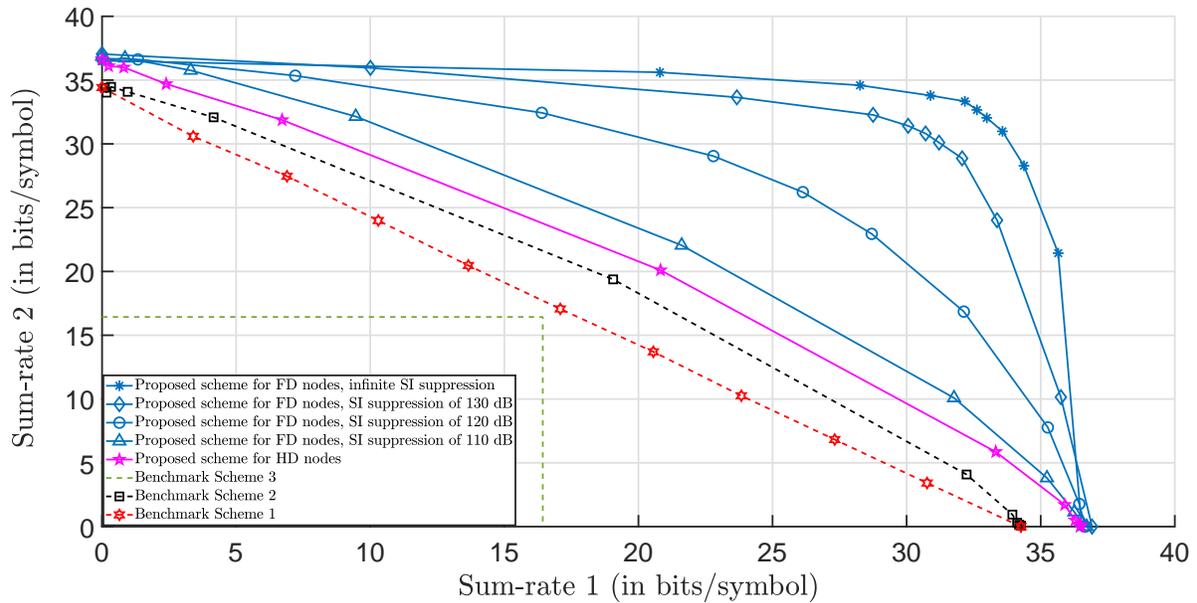}
\caption{Sum-rate 1 vs. sum-rate 2 of the proposed schemes and the benchmark schemes for $P$=20 dBm, $\rho$=1000 $m$.}
\label{Rate_region}
\end{figure}

 In Fig.~\ref{Rate_region}, we show the rate region achieved using the optimal centralized D-TDD scheme  for two different group of nodes, where all the nodes that belong in each group have the same values of $\mu$. Let $\mu_1$ be assigned to the first group and $\mu_2$ to the second group of nodes. By varying the value of $\mu_1$ from zero to one, and setting $\mu_2=1-\mu_1$, as well as aggregating  the achieved rates for each group we can get the rate region of the network of the two groups. In this example, the transmit power is fixed to  $P$=20 dBm and the area dimension is 1000$\times$1000  $m^2$. As shown in Fig.~\ref{Rate_region}, the proposed scheme with HD nodes has more than 15$\%$ improvement in the rate region area compared to the benchmark schemes. More importantly, the proposed scheme for FD nodes with SI suppression of 110 dB  performs approximately four times better then  Benchmark Scheme 3, in addition to  outperforming the other benchmark schemes as well, which is a huge gain and a promising result for using FD nodes.

\begin{figure}[t]
\vspace*{-2mm}
\centering\includegraphics[width=6.7in]{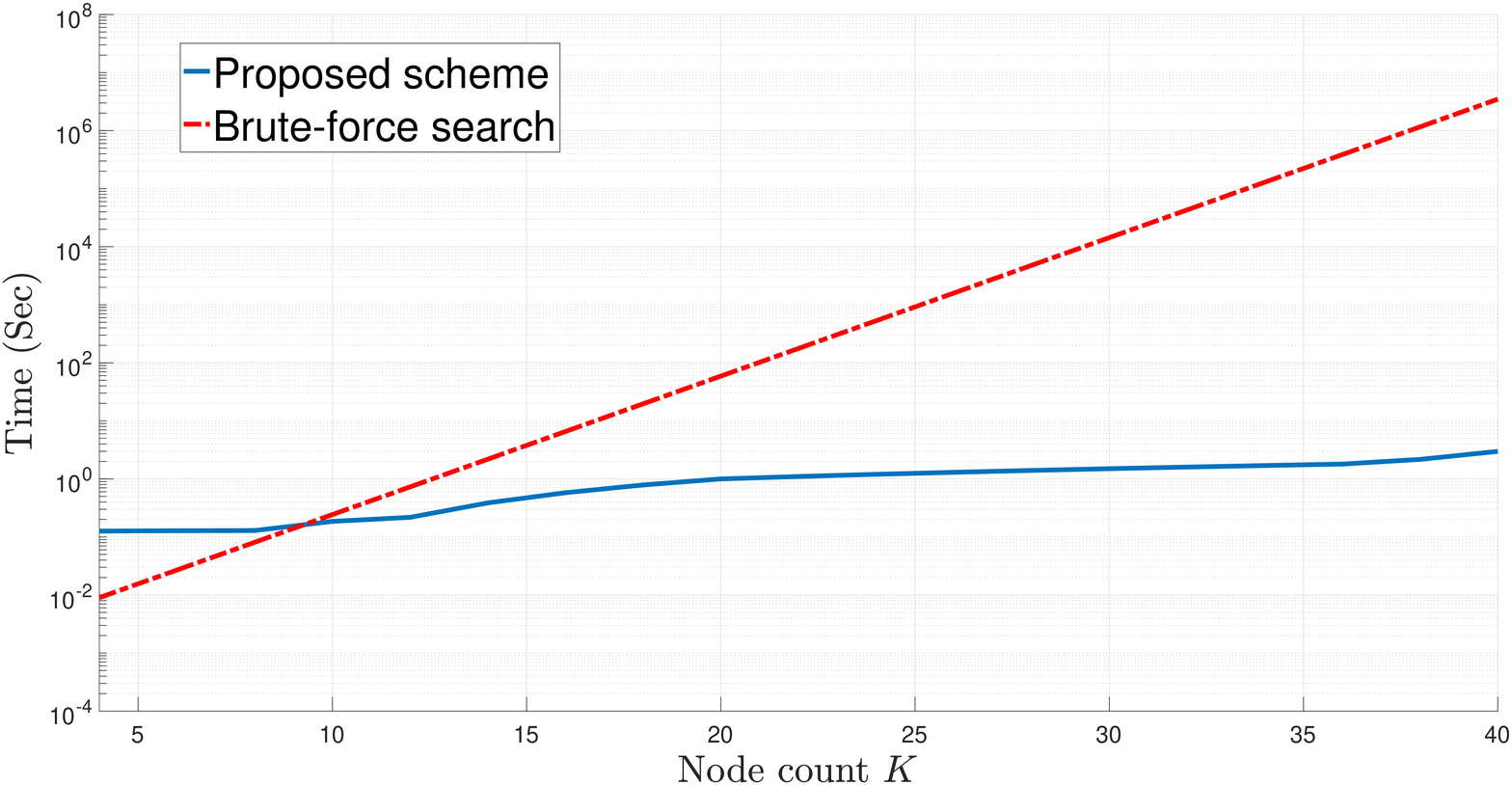}
\caption{Complexity vs. node count of the proposed schemes and the benchmark scheme for $P$=20 dBm and $\rho$=1000 $m$.}
\label{Time_Node}
\end{figure}

In Fig.~\ref{Time_Node}, we present the total time required by the optimal algorithm presented in Algorithm~\ref{Lyp_algorithm} to obtain the solution as a function of the number of nodes in the network. For comparison purpose, we also present the total time required by a general brute-force search algorithm to search over all the possible solutions in order to to find the optimal one. To this end, we set the power at the nodes to $P=20$ dBm, and the area to 1000$\times$1000  $m^2$. As it can be seen from Fig.~\ref{Time_Node}, the brute-force search algorithm's computation time increases exponentially, however, the computation time with the proposed algorithm increases linearly.

\begin{figure}[t]
\vspace*{-2mm}
\centering\includegraphics[width=6.7in,height=4in]{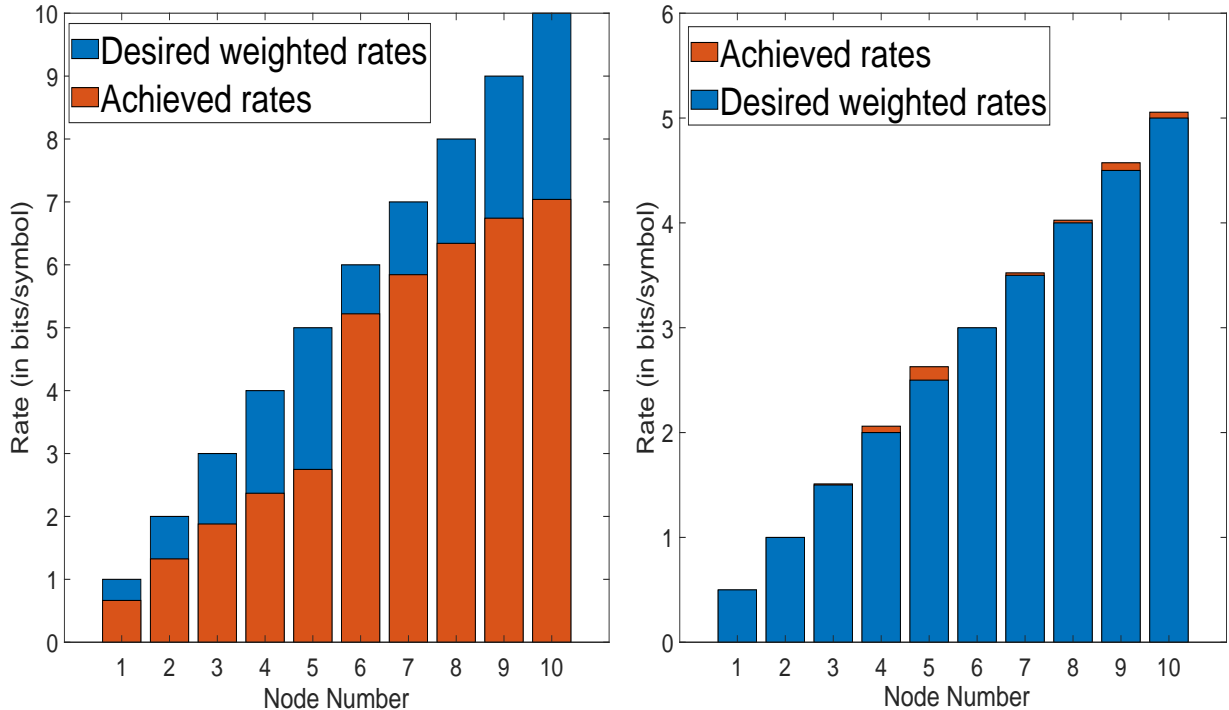}
\caption{Rate vs. node number of the proposed scheme applying rate allocation scheme with under-capacity data demand (right), and with over-capacity data demand (left),   $P$=20 dBm. SI=110 dB, and $\rho$=1000 $m$.}
\label{Fainess_Node}
\end{figure}

In Fig.~\ref{Fainess_Node}, we illustrate the rate  achieved using the proposed scheme applying the rate allocation scheme for $N=10$, as a function of node number index. Moreover, we assume that the transmit power is fixed to $P=20$ dBm, the dimension is  $\rho$=1000 $m$, and the SI suppression is 110 dB. We have investigated two cases where in both cases  the users have same priority, i.e., $\alpha_k=0.1, \forall k$.  However, in one case data demand by users (right plot) is set to  $\tau_k=\frac{k}{2}, \forall k$, and in the other the data demand by users (left plot) is set to $\tau_k=k, \forall k$. As can be seen in the right plot of the Fig.~\ref{Fainess_Node}, the rate allocation scheme is able to successfully answer the data demanded by users. However, in the case of the left plot of Fig.~\ref{Fainess_Node}, the  rate allocation scheme 
was not able to answer the rates demand of the nodes due to capacity limits. Regardless, it successfully managed to hold the average received rates as close as possible to the demanded rates.

\section{Conclusion}\label{Sec-Conc}
In this paper, we devised  the optimal  centralized D-TDD scheme for a wireless network comprised of $K$ FD or   HD  nodes, which maximizes the rate region of the network. The proposed centralized D-TDD scheme makes an optimal decision of which node should receive, transmit, simultaneously receive and transmit, or be silent in each time slot. In addition, we proposed a fairness scheme that allocates data rates to the nodes according to the user data demands. We have shown that the proposed optimal  centralized D-TDD scheme has significant gains over  existing  centralized D-TDD schemes.

\appendix 
\subsection{Proof of Theorem~\ref{theo_1}}\label{app_PR_Local}
The signal received  at node $k$ is given by
\begin{align}\label{Eqe_SINR_1}
y_k(i)&= r_k(i)  \left( \sum_{v\in \mathcal{K}_d} [t_v(i)+f_v(i)] \sqrt{P g_{v,k}(i)} \mathbf{s}_v(i)+\sum_{v\in \mathcal{K}_u}  [t_v(i)+f_v(i)] \sqrt{P g_{v,k}(i)} \mathbf{s}_v(i) \right)\\
&+f_k(i) \left( \sum_{v\in \mathcal{K}_d} [t_v(i)+f_v(i)] \sqrt{P g_{v,k}(i)} \mathbf{s}_v(i)+\sum_{v\in \mathcal{K}_u}  [t_v(i)+f_v(i)] \sqrt{P g_{v,k}(i)} \mathbf{s}_v(i) \right. 
\Bigg) + n_k(i)\nonumber
\end{align} 
where $\mathcal{K}_d$ and $\mathcal{K}_u$  are the sets of desired and undesired nodes, respectively and $\mathbf{s}_v(i)$ is the transmitted codeword from node $v$. Assuming the transmission rates of the desired nodes are adjusted such that the receiving node can perform successive interference cancellation of the desired codewords, the rate received  at node $k$ from all the desired nodes is given by
\begin{align}\label{Eqe_SINR_3dsfg}
R_k(i)=  \log_2\left(1+ \frac{[r_k(i)+f_k(i)]\sum_{v\in \mathcal{K}_d} [t_v(i)+f_v(i)]  P g_{v,k}(i)} {\sigma_k^2+[r_k(i)+f_k(i)]  \sum_{v\in \mathcal{K}_u} [t_v(i)+f_v(i)]  P g_{v,k}(i)   }\right),
\end{align} 
which can be simplified to
\begin{align}\label{Eqe_SINR_3}
R_k(i)=  \log_2\left(1+ \frac{[r_k(i)+f_k(i)]\sum_{v\in \mathcal{K}_d} [t_v(i)+f_v(i)]  P g_{v,k}(i)} {\sigma_k^2+  \sum_{v\in \mathcal{K}_u} [t_v(i)+f_v(i)]  P g_{v,k}(i)   }\right).
\end{align}
By substituting $\sum_{v\in \mathcal{K}_d}g_{v,k}(i)=\left [\mathbf{t}(i)+\mathbf{f}(i) \right ]\mathbf{d}_k^{\intercal}(i)$ and  $\sum_{v\in \mathcal{K}_u}g_{v,k}(i)=\left [\mathbf{t}(i)+\mathbf{f}(i) \right ]  \mathbf{i}_k^{\intercal}(i)$ into (\ref{Eqe_SINR_3}), and  assuming that $\sigma_k^2=1$, we obtain the  rate   $R_k(i)$ as in (\ref{Eqe_SINR_3as}). This completes the proof.

\subsection{Proof of Theorem~\ref{theo_2}}\label{app_PR_Outage_L} 
 Using the vector $\mathbf{t}(i)$ and the matrix $\mathbf{ Q}$, we can obtain the other vectors, $\mathbf{r}(i)$, $\mathbf{f}(i)$, and $\mathbf{s}(i)$. Specially, if  $t_x(i)=0$, $t_k(i)=1$, and the $(x,k)$ element of $\mathbf{ Q}$ is equal to one, then, $r_x(i)=1$.  If  $t_x(i)=0$, $t_k(i)=0$, and $(x,k)$ element of $\mathbf{ Q}$ is equal to one, then, $s_x(i)=1$.   If  $t_x(i)=1$, $t_k(i)=1$, and $(x,k)$ element of $\mathbf{ Q}$ is equal to one, then, $f_x(i)=1$ and we set $t_x(i)=0$. Finally, if $t_x(i)=1$, $t_k(i)=0$, and $(x,k)$ element of $\mathbf{ Q}$ is equal to one, then, $t_x(i)$ is given by (\ref{eq_scheme_D-TDD}).

Since the values of $\mathbf{t}(i)$ are sufficient, we simplify the optimization problem in (\ref{eq_max_1a}) as
 \begin{align}
& {\underset{\mathbf{t}(i), \forall i} {  \textrm{Maximize:} }}\; 
  \lim_{T\to\infty}\frac{1}{T} \sum_{i=1}^T \sum_{k=1}^N     \mu_k \log_2 \left(1+\frac{P\,\mathbf{t}(i)  \mathbf{d}_k^{\intercal}(i)}{1+ P\, \mathbf{t}(i)  \mathbf{i}_k^{\intercal}(i)} \right) \nonumber\\
&{\rm{Subject\;\;  to \; :}}  \nonumber\\
&\qquad\qquad{\rm C1:}\;  t_v(i)\in\{0,1\}, \; \forall v.
\label{eq_max_1a_app}
\end{align}

To obtain the solution of (\ref{eq_max_1a_app}), we first transform the non-convex objective function in   (\ref{eq_max_1a_app}) into an equivalent objective function. To this end, let us define  $A_{k}(i)$ and $B_{k}(i)$ as the numerator and denominator values to simplify the notation, where
\begin{align}
A_{k}(i)&= P  \mathbf{t}(i)  \mathbf{d}_k^{\intercal}(i),\label{eq_7aa}\\
B_{k}(i)&= 1+P  \mathbf{t}(i)  \mathbf{i}_k^{\intercal}(i).\label{eq_7bb}
\end{align}

 Now, we relax constraint C1 in (\ref{eq_max_1a}) such that $0\leq  t_v(i)\leq 1$, $0\leq  r_v(i)\leq 1$, and $0\leq  t_v(i)+r_v(i)\leq 1$, $\forall v$, respectively, and rewrite (\ref{eq_max_1a})  as
 
 \begin{align}
& {\underset{\mathbf{t}(i),\forall i} {  \textrm{Maximize:} }}\; 
  \lim_{T\to\infty}\frac{1}{T}\sum_{i=1}^T \sum_{k=1}^N \mu_k \log_2 \left(1+\frac{{A}_{k}(i)}{{B}_{k}(i)} \right) \nonumber\\
&{\rm{Subject\;\;  to \; :}}  \nonumber\\
&\qquad\qquad{\rm C1:}\;  0\leq  t_v(i)\leq 1, \; \forall v.
\label{eq_max_app_2}
\end{align}

Using Proposition 1 in \cite{7381695}, we transform the objective function in (\ref{eq_max_app_2}) into an equivalent  form as
 \begin{align}
& {\underset{\mathbf{t}(i), \mathbf{w}(i), \forall i} {  \textrm{Maximize:} }}\; 
  \lim_{T\to\infty}\frac{1}{T}\sum_{i=1}^T \sum_{k=1}^N \mu_k \log_2 \left( \frac{|w_k(i)|^2}{|\sqrt{{A}_{k}(i)}-w_k(i)|^2+{B}_{k}(i)}  \right) \nonumber\\
&{\rm{Subject\;\;  to \; :}}  \nonumber\\
&\qquad\qquad{\rm C1:}\;  0\leq  t_v(i)\leq 1, \; \forall v,
\label{eq_max_app_3}
\end{align} 
where the vector $\mathbf{w}(i)=\{w_1(i),w_2(i),...,w_N(i)\}$ is a scaling factor vector, given by
 \begin{align}
w_k(i)=\frac{{A}_{k}(i)+{B}_{k}(i)}{\sqrt{{A}_{k}(i)}}.
\label{eq_max_app_4}
\end{align} 

It has been shown in Proposition 1 in \cite{7381695} that the optimization problem in (\ref{eq_max_app_3}) 
is equivalent to the optimization problem in (\ref{eq_max_app_2}) when the scaling factor  $\mathbf{w}(i)$ is optimized using (\ref{eq_max_app_4}), i.e., both (\ref{eq_max_app_3}) and (\ref{eq_max_app_2}) have the same global solution when $\mathbf{w}(i)=\mathbf{W^{opt}}(i)$ is selected optimally. When $\mathbf{w}(i)$ is obtained from (\ref{eq_max_app_4}), the optimization problem in (\ref{eq_max_app_3}) can be written as an optimization of $\mathbf{t}(i)$ as
 \begin{align}
& {\underset{\mathbf{t}(i), \forall i} {  \textrm{Maximize:} }}\; 
  \lim_{T\to\infty}\frac{1}{T}\sum_{i=1}^T  \sum_{k=1}^N \mu_k\Bigg( \log_2 \left( |w_{k}^{opt}(i)|^2  \right)-\log_2 \left( |\sqrt{{A}_{k}(i)}-w_{k}^{opt}(i)|^2+{B}_{k}(i)  \right)\Bigg) \nonumber\\
&{\rm{Subject\;\;  to \; :}}  \nonumber\\
&\qquad\qquad{\rm C1:}\;  0\leq  t_v(i)\leq 1, \; \forall v.
\label{eq_max_app_5}
\end{align} 
However, the optimization problem in (\ref{eq_max_app_5}) is still non-convex \cite{J_Conway_b}. Hence, we define an additional scaling factors vector, $\mathbf{l}(i)=\{l_1(i),l_2(i),...,l_N(i)\}$, and rewrite (\ref{eq_max_app_5}) as
 \begin{align}
& {\underset{\mathbf{t}(i),\mathbf{l}(i), \forall i} {  \textrm{Maximize:} }}\; 
  \lim_{T\to\infty}\frac{1}{T}\sum_{i=1}^T  \sum_{k=1}^N \mu_k \Bigg(\log_2 \left( |w_{k}^{opt}(i)|^2  \right)+\log_2(l_k(i))\nonumber\\
  &\qquad\qquad-\frac{l_k(i)}{\ln{2}}  \left( |\sqrt{{A}_{k}(i)}-w_{k}^{opt}(i)|^2+{B}_{k}(i)  \right) \Bigg) \nonumber\\
&{\rm{Subject\;\;  to \; :}}  \nonumber\\
&\qquad\qquad{\rm C1:}\;  0\leq  t_v(i)\leq 1, \; \forall v,
\label{eq_max_app_6}
\end{align} 
where clearly (\ref{eq_max_app_6}) is a concave function of  $\mathbf{l}(i)$. Furthermore, the optimum $\mathbf{l}(i)$ can be calculated by taking  the derivative from the objective function in  (\ref{eq_max_app_6}) with respect to $\mathbf{l}(i)$ and then setting the result to zero,
which results in
 \begin{align}
l_k(i)=\frac{1}{\left( |\sqrt{{A}_{k}(i)}-w_{k}^{opt}(i)|^2+{B}_{k}(i)  \right)}.
\label{eq_max_app_7}
\end{align}

The  optimization problem in (\ref{eq_max_app_6}) has the same solution as the main optimization problem in (\ref{eq_max_app_5}), when $\mathbf{w}(i)$  and $\mathbf{l}(i)$ are chosen using (\ref{eq_max_app_4}) and (\ref{eq_max_app_7}), respectively. As a result, our problem now is 
 \begin{align}
& {\underset{\mathbf{t}(i), \forall i} {  \textrm{Maximize:} }}\; 
  \lim_{T\to\infty}\frac{1}{T}\sum_{i=1}^T  \sum_{k=1}^N \mu_k \Bigg(\log_2 \left( |w_{k}^{opt}(i)|^2  \right)+\log_2(l_k^{opt}(i))\nonumber\\
  &\qquad\qquad-\frac{l_k^{opt}(i)}{\ln{2}}  \left( |\sqrt{{A}_{k}(i)}-w_{k}^{opt}(i)|^2+{B}_{k}(i)  \right) \Bigg) \nonumber\\
&{\rm{Subject\;\;  to \; :}}  \nonumber\\
&\qquad\qquad{\rm C1:}\;  0\leq  t_v(i)\leq 1, \; \forall v.
\label{eq_max_app_8}
\end{align} 

We now use the Lagrangian to solve (\ref{eq_max_app_8}). Thereby, we obtain
\begin{align}\label{eq_op_Lr1}
{\cal L} =&\lim_{T\to\infty}\frac{1}{T}\sum_{i=1}^T  \sum_{k=1}^N  \frac{\mu_k l_k^{opt}(i)}{\ln{2}}  \left( |\sqrt{{A}_{k}(i)}-w_{k}^{opt}(i)|^2+{B}_{k}(i)  \right)\nonumber\\
&-\sum_{v=1}^N { {\lambda _1^v(i)}  t_v(i) }-\sum_{v=1}^N {\lambda _2^v(i)\left(1 -  t_v(i)\right) },
\end{align}
where $\lambda _1^v(i)\geq0$ and $\lambda _2^v(i)\geq0$, $\forall v$, are the Lagrangian multipliers. By differentiating ${\cal L}$ in (\ref{eq_op_Lr1}) with respect to $t_x(i)$, $\forall x$, we obtain
\begin{align}\label{eq_op_Lr2}
\frac{d{\cal L} }{d t_x(i)} =& \sum_{k=1}^N  \frac{P \mu_k l_k^{opt}(i)}{\ln{2}}  \left[  d_{x,k}(i) \left (1-\frac{w_{k}^{opt}(i)}{\sqrt{{A}_{k}(i)}} \right)+i_{x,k}(i)  \right]-\lambda _1^x(i) +\lambda _2^x(i).
\end{align}
Finally, equivalenting the results in (\ref{eq_op_Lr2})  to zero, $\frac{d{\cal L} }{d t_x(i)}=0$, gives us the necessary equations to acquire  optimum $t_x(i)$, $\forall x$, as 
\begin{align}\label{eq_op_Lr3}
\sum_{k=1}^N  \frac{P \mu_k l_k^{opt}(i)}{\ln{2}}  \left[  d_{x,k}(i) \left (1-\frac{w_{k}^{opt}(i)}{\sqrt{{A}_{k}(i)}} \right)+i_{x,k}(i)  \right]-\lambda _1^x(i) +\lambda _2^x(i)=0, \forall i .
\end{align}

In order to find the condition for specifying the value of one or zero to each $t_x(i)$, $\forall x$, we set $t_x(i)=0$ in  (\ref{eq_op_Lr3}) which leads $\lambda _2^x(i)=0$ (by
complementary slackness in KKT condition), as a result the condition for choosing $t_x(i)=0$ is acquired as 

\begin{align}\label{eq_op_Lr4}
\sum_{k=1}^N  \frac{P \mu_k l_k^{opt}(i)}{\ln{2}}  \left[  d_{x,k}(i) \left (1-\frac{w_{k}^{opt}(i)}{\sqrt{P\sum_{v=1,v \ne x}^N {t_v(i) d_{v,k}(i)} }} \right)+i_{x,k}(i)  \right]=\lambda _1^x(i), \forall v,i .
\end{align}

By knowing that $\lambda _1^x(i)\geq0$, we obtain the optimal state selection
scheme in Theorem~\ref{theo_2}. This completes the proof.

\subsection{Proof of Theorem~\ref{theo_2}}\label{app_PR_Outage_L3} 
The main diagonal elements of $\mathbf{G}(i)$ model the SI channel of each node.  Hence, by setting the values of the main diagonal of  $\mathbf{G}(i)$ to infinite, we will make the simultaneous reception and transmissions  for the FD nodes  impossible to be selected and thereby make the FD nodes into HD nodes. As a result, in the proposed centralized D-TDD scheme  in Algorithm~\ref{Lyp_algorithm}, the nodes  will either be transmitting, receiving, or be silent. Hence, the proposed  scheme in Algorithm~\ref{Lyp_algorithm} is the optimal centralized D-TDD scheme for a wireless network comprised of HD nodes when the main diagonal of the  $\mathbf{G}(i)$ are set to infinity.

\bibliography{litdab}
\bibliographystyle{IEEEtran}
\end{document}